\documentclass[reprint,pra,aps,notitlepage,nofootinbib,superscriptaddress]{revtex4-2}
\usepackage{amsmath}
\usepackage{amssymb}
\usepackage{bm,braket}
\usepackage{float} 
\usepackage{xurl}
\usepackage{hyperref}
\usepackage{multirow}
\hypersetup{
     colorlinks   = true,
     citecolor    = blue
}
\def\be{\begin{equation}}
\def\ee{\end{equation}}
\def\bea{\begin{eqnarray}}
\def\eea{\end{eqnarray}}
\def\bes{\begin{eqnarray}}
\def\ees{\end{eqnarray}}
\def\bi{\begin{itemize}}
	\def\ei{\end{itemize}} 
	
\usepackage{amsthm}

\theoremstyle{definition}

\usepackage{tikz}
\usetikzlibrary{braids}

\begin{document}
\title{Simulating Neutron Scattering on an Analog Quantum Processor  } 
\author{Nora Bauer}
\email{nbauer1@vols.utk.edu}
\affiliation{Department of Physics and Astronomy,  The University of Tennessee, Knoxville, TN 37996-1200, USA}

\author{Victor Ale}
\email{vale@vols.utk.edu}
\affiliation{Department of Physics and Astronomy,  The University of Tennessee, Knoxville, TN 37996-1200, USA}

\author{Pontus Laurell}
\email{plaurell@missouri.edu}
\affiliation{Department of Physics and Astronomy,  The University of Tennessee, Knoxville, TN 37996-1200, USA}
\affiliation{Department of Physics and Astronomy,  The University of Missouri, Columbia, MO 65211, USA}

\author{Serena Huang}
\email{serenaqhuang@gmail.com}
\affiliation{Department of Computer Science,  The University of Maryland, MD 20742-5025, USA}

\author{Seth Watabe}
\email{sethwatabe@gmail.com}
\affiliation{Department of Computer Science,  Rensselaer Polytechnic Institute, NY 12180, USA}

\author{David Alan Tennant}
\email{dtennant@utk.edu}
\affiliation{Department of Physics and Astronomy,  The University of Tennessee, Knoxville, TN 37996-1200, USA}
\affiliation{Department of Materials Science and Engineering,  The University of Tennessee, Knoxville, TN 37996-2100, USA}
\author{George Siopsis}
\email{siopsis@tennessee.edu}
\affiliation{Department of Physics and Astronomy,  The University of Tennessee, Knoxville, TN 37996-1200, USA}	 

\date{\today}
\begin{abstract}
Neutron scattering characterization of materials allows for the study of entanglement and microscopic structure, but is inefficient to simulate classically for comparison to theoretical models and predictions. However, quantum processors, notably analog quantum simulators, have the potential to offer an unprecedented, efficient method of Hamiltonian simulation by evolving a state in real time to compute phase transitions, dynamical properties, and entanglement witnesses. Here, we present a method for simulating neutron scattering on QuEra's Aquila processor by measuring the dynamic structure factor (DSF) for the prototypical example of the critical transverse field Ising chain, and propose a method for error mitigation. We provide numerical simulations and experimental results for the performance of the procedure on the hardware, up to a chain of length $L=25$. Additionally, the DSF result is used to compute the quantum Fisher information (QFI) density, where we confirm bipartite entanglement in the system experimentally. 
\end{abstract}
\maketitle 
\twocolumngrid 
\section{Introduction}

Entanglement is a fundamental resource in quantum technologies, and a central property in many-body physics. A key challenge in experiment is the quantification of multipartite entanglement. To this end quantum Fisher information (QFI) is of emerging significance as it serves as a witness for multipartite entanglement and can be measured through dynamic susceptibilities \cite{Hauke_2016} , which are experimentally accessible through spectroscopy in cold atomic gases and condensed matter systems. This links entanglement with many-body correlations in response functions, especially near quantum phase transitions, where QFI can exhibit universality, identifying strongly entangled phases. This framework is applicable to lattice models at thermal equilibrium, such as quantum Ising models, with potential signatures observable in optical-lattice experiments and strongly correlated materials \cite{Hauke_2016, LaurellEntanglementReview}.

The key experimental quantity in such experiments is the dynamical structure factor (DSF), which reflects the system's response to perturbations and  reveals quantum dynamics and static order. In the case of quantum spin systems, neutron scattering is particularly valuable as it directly probes correlations between spins through the dynamical spin structure factor, which can be measured to high accuracy. A number of numerical methods have been introduced for its computation on classical computers including (i) the Lanczos method \cite{Dagotto_1994}, which is limited to small system sizes, (ii) quantum Monte Carlo (QMC) methods \cite{Jarrell_1996}, which involve an ill-defined analytic continuation from imaginary to real time and often are inapplicable to systems of interest due to sign problems \cite{Troyer_2005}, and (iii) DMRG \cite{White_1992} and tensor network methods, which are particularly suitable to low-dimensional systems such as the quantum Ising chain, but suffer in performance for highly entangled states. Quantum simulators, in contrast, inherently have the potential to handle both large systems and highly entangled states \cite{Baez_2020}.

DMRG and its reformulations in the matrix product state (MPS) language \cite{Schollwock_2011} are traditionally used for ground-state and finite-temperature calculations in quasi-one-dimensional systems, but can be extended to compute dynamical properties via three key methods: (i) the continued fraction expansion \cite{Dagotto_1994}, which provides quick but less accurate results at higher frequencies, (ii) the correction vector method, which computes dynamical correlation functions at specific frequencies \cite{DCF1999, ORNL2016DSF} such that the computational cost grows with the number of evaluated frequencies, and (iii) approaches involving evolution in real time \cite{Schollwock_2011, Paeckel_2019}, which have to contend with entanglement growth with time and often utilize extrapolation to reach very low frequencies. The latter two are highly useful for studying a range of low-dimensional systems, including Heisenberg, t-J, and Hubbard models. These approaches allow us to probe the evolution of entanglement and phase transitions in quantum systems, providing crucial insight into strongly correlated materials and quantum technologies. 

However, the drawbacks of each method effectively limit the scale and resolution that can be reached. These difficulties can be addressed by quantum simulators, which naturally track the time evolution of such systems. Quantum simulations using cold atom arrays, trapped ions, and superconducting qubits can simulate dynamical structure factors, enabling observation of correlated quantum matter's dynamics, even in the presence of noise. This approach holds promise for achieving results beyond classical computational limits for larger system sizes, relevant for neutron scattering experiments in condensed matter \cite{Baez_2020}.

In the context of quantum simulations, the Lanczos algorithm combined with matrix product states (MPS) offers a powerful tool for calculating dynamical correlation functions. This technique improves the accuracy and convergence of spectral weights in models like the spin-1/2 antiferromagnetic Heisenberg chain \cite{SpinChain2012}. By addressing issues such as the ghost problem of Lanczos methods, the MPS-based approach delivers highly accurate results, even when compared to exact methods like the Bethe Ansatz. This improvement is critical for simulating dynamical properties such as spin excitations, which are needed for undertaking interpretation and quantitative analysis of neutron experiments.

Out-of-time-order correlators (OTOCs) have emerged as a promising tool for learning about quantum systems, particularly when conventional observables fail due to rapid decay in strongly interacting systems. OTOCs provide valuable information at large times and distances, making them effective for characterizing systems with weak or hidden interactions. By offering an exponential advantage in learning tasks like Clifford tomography, OTOCs open up new pathways for probing complex quantum systems \cite{PhysRevResearch.5.043284}.

MPS have become a standard for simulating one-dimensional quantum systems. Various methods, such as time-evolving block decimation and the Krylov method, have been developed to handle time evolution within MPS frameworks \cite{Paeckel_2019}. These methods are crucial for studying time-dependent phenomena, such as phase transitions and spin dynamics, in quantum systems. The ability to evolve MPS accurately allows us to simulate neutron scattering experiments, where time-dependent correlation functions play a key role in interpreting the material's quantum behavior.

Together, these quantum simulation methods push the boundaries of what can be observed in neutron scattering experiments, allowing for the exploration of larger, more complex quantum systems than classical methods permit.







Rydberg atom simulators have gained attention due to their ability to explore complex quantum systems. These simulators exploit the high tunability and strong interactions between Rydberg atoms to simulate various quantum phenomena, including optimization and quantum error correction.

Aquila, QuEra’s 256-qubit quantum simulator, operates on a field-programmable qubit array, enabling programmable quantum dynamics across neutral-atom qubits \cite{wurtz2023aquila}. It is an analog Hamiltonian simulator, capable of tackling complex quantum states with configurable architectures. Its applications range from single-qubit dynamics to combinatorial optimization problems, like the native maximum independent set problem, making it a versatile tool for exploring neutral-atom quantum computing.  The processor can also be used for state preparation and optimization of non-native problems, such as MaxCut, which are relevant for broad classes of optimization tasks such as power grid optimization \cite{bauer2024solvingpowergridoptimization}. 

Reconfigurable Rydberg atom arrays offer significant potential for quantum error correction. Recent advancements include the development of a logical quantum processor that supports up to 280 physical qubits, capable of operating with high-fidelity two-qubit gates and arbitrary connectivity \cite{Bluvstein2024}. This system has demonstrated fault-tolerant generation of logical Greenberger-Horne-Zeilinger (GHZ) states and improved algorithmic performance through logical encoding, charting a path toward large-scale quantum computing.

Rydberg atom arrays are also instrumental in solving NP-hard combinatorial optimization problems. Ebadi et al.\ \cite{ebadi2022quantum} demonstrated the use of Rydberg blockade to encode and solve the maximum independent set problem on programmable graphs. The ability to explore non-planar graphs and test variational algorithms reveals the power of Rydberg simulators in quantum optimization, pushing the boundaries of what quantum systems can achieve in hard computational tasks.

Rydberg quantum wires, which exploit auxiliary atoms, can solve complex combinatorial problems on three-dimensional atom arrays. This approach overcomes the intrinsic limitations of two-dimensional arrays by encoding high-degree vertices and constructing solutions for non-planar and high-degree graphs. By leveraging many-body entanglement, this protocol represents a step towards demonstrating quantum advantage in solving combinatorial optimization problems \cite{Kim2022}.

Aquila also supports the generation of complex quantum states, including Bell state entanglement and $\mathbb{Z}_2$ states with defects \cite{balewski2024engineering}. By optimizing driving fields and readout error mitigation techniques, these experiments showcase Aquila's precision and alignment with theoretical models. Such capabilities not only highlight Aquila's performance in handling advanced quantum simulations but also open new research directions in quantum information processing.






Quantum simulators, particularly those based on Rydberg atom arrays, offer a powerful platform for investigating the complex dynamics of quantum systems. These arrays enable the precise control and manipulation of individual atoms, allowing us to simulate a wide range of quantum phenomena, including antiferromagnetic spin states, topological phases, and many-body dynamics.

One significant application of quantum simulators is the study of antiferromagnetic spin states with domain wall defects \cite{balewski2024engineering,q-ctrl-2}. In these systems, domain walls represent boundaries between regions with different magnetic orders and play a critical role in exploring quantum phase transitions and collective excitations. Rydberg arrays, with their ability to manipulate individual spins, allow for real-time investigation of the formation and dynamics of these domain walls, providing valuable insights into topological excitations and magnetic systems.

Another fascinating area of study is the realization of topological spin liquids on Rydberg atoms \cite{Semeghini2021}. These exotic quantum phases lack local order but exhibit long-range entanglement and topological order, making them distinct from conventional magnetic phases. Rydberg atom arrays are ideal for simulating the frustration and long-range interactions necessary for these spin liquids, advancing the understanding of highly entangled quantum states, which hold potential applications in fault-tolerant quantum computing.

The ability to control quantum many-body dynamics in driven Rydberg arrays is another significant advancement \cite{Bluvstein2021}. By applying time-dependent driving fields, we can control interactions between atoms, leading to phenomena such as dynamical phase transitions and out-of-equilibrium states. These arrays provide a means to study quantum quenches and explore the non-equilibrium behavior of strongly correlated systems, helping to uncover phenomena that are difficult or impossible to simulate on classical computers.

Rydberg atom arrays also provide a platform for simulating quantum Ising models in tunable two-dimensional geometries \cite{Labuhn_2016}, allowing us to explore quantum magnetism, phase transitions, and quantum criticality. These programmable arrays enable the precise control of interactions between atoms, making them an ideal tool for investigating complex spin correlations and critical phenomena, with implications for quantum materials and technologies like quantum annealing.

Additionally, quantum simulation of 2D antiferromagnets with hundreds of Rydberg atoms has become a reality \cite{Scholl_2021}, providing an analog for real antiferromagnetic materials. These simulations allow us to study how quantum fluctuations affect magnetic order in two dimensions, shedding light on emergent quantum phases and potential quantum phase transitions. This capability is critical for understanding complex materials such as high-temperature superconductors.

Beyond conventional quantum phases, non-Abelian Floquet spin liquids can be realized using periodic driving in Rydberg systems \cite{Lukin2023}. These spin liquids host non-Abelian anyons, particles with unique statistics that could be harnessed for topological quantum computation. Rydberg arrays provide the ability to implement Floquet protocols that stabilize these exotic phases, opening new avenues for quantum error correction and robust quantum information processing.

Finally, probing many-body dynamics on large quantum simulators \cite{Bernien_2017} offers a testbed for exploring collective quantum behavior. These simulators allow us to investigate phenomena such as thermalization, many-body localization, and dynamical phase transitions, providing insight into how collective behavior emerges in large quantum systems.

Thus, Rydberg atom arrays have advanced the field of quantum simulation, offering a versatile and powerful tool for studying a wide range of quantum phenomena. From antiferromagnetic spin states to topological phases and many-body dynamics, these simulators are pushing the boundaries of what can be explored and understood in the quantum world, with profound implications for quantum computing, material science, and fundamental physics.

Here, we present a method for simulating neutron scattering experiments using QuEra’s Aquila analog quantum processor, particularly focusing on the transverse field Ising  (TFI) model. The ground state of the critical TFI model was prepared using an adiabatic Ansatz optimized for the Aquila processor. Two approaches, Quantum Approximate Optimization Algorithm (QAOA) and an adiabatic Ansatz, were compared, with the latter providing higher fidelity.
The DSF was computed by measuring the retarded Green's function in time and space using the Aquila processor. The measurement involved applying a local spin operator and evolving the system with the Ising Hamiltonian.
We introduced a noise model to simulate hardware errors, including State Preparation and Measurement (SPAM) errors, laser-induced noise, and effective thermal noise. Error mitigation techniques, such as SPAM correction using confusion matrices, were applied to reduce noise impact.
Systematic errors from the state preparation and time evolution were evaluated. The main source of error was found in the state preparation process, while the next-nearest neighbor interactions had minimal effect.
For chain lengths up to 
$L=25$, experimental results from Aquila closely matched theoretical predictions after error mitigation. The mitigated results displayed significant improvement over raw experimental data.
The QFI density was computed as an entanglement witness, showing at least bipartite entanglement in the system after error mitigation. For larger systems, experimental results were compared with tensor network simulations using DMRG, showing close agreement.

Our discussion is organized as follows. In Section \ref{sec:method}, we describe the adiabatic-type Ansatz to prepare the ground state of the critical TFI model and the procedure to measure the DSF. In Section \ref{sec:error}, we detail our noise model used in numerical calculations and propose an error mitigation procedure for measuring the DSF experimentally. In Section \ref{sec:numerical}, we present numerical results studying the impact of systematic noise from approximations made in the procedure and hardware noise estimated using the noise model. In Section \ref{sec:experimental}, we give experimental results for the DSF measured on the Aquila processor and compute the QFI entanglement measure. Finally, in Section \ref{sec:conclusion}, we summarize our results and discuss future directions. 









\section{Method}\label{sec:method} 
Here we describe the method for preparing the ground state of the critical TFI chain and measuring the DSF on the Aquila processor. For the ground state preparation, we use an adiabatic-type Ansatz that is optimized over a set of hyperparameters. Then, the DSF is computed by measuring the retarded space-time Green's function as a function of time, where the time evolution operator is applied directly on the analog processor. We also discuss the systematic errors that these procedures introduce. 

The Aquila processor from QuEra \cite{wurtz2023aquila} allows time  evolution of a two-dimensional array of Rydberg atoms governed by the time-dependent Hamiltonian
\bea H(t) &=& \frac{1}{2} \sum_{i=1}^n \Delta_{i} (t)Z_i + \frac{\Omega(t)}{2} \sum_{i=1}^n [ X_i \cos \phi(t) - Y_i \sin\phi(t) ]  \nonumber\\ && + \frac{C_6}{4} \sum_{i<j} \frac{(Z_i -\mathbb{I}) (Z_j - \mathbb{I})}{[(x_i-x_j)^2 + (y_i-y_j)^2]^3}  \label{eq:aquila} \eea
where $C_6$ is the coupling constant of the van der Waals force. The (fixed) positions of the atoms, $(x_i,y_i)$, as well as the time-dependent functions $\Delta_{i} (t), \Omega(t), \phi(t)$, can be chosen at will.

We restrict our focus to the transverse field Ising model with Hamiltonian 
\be H=J\left(\sum_{\langle i,j\rangle}Z_iZ_j+g\sum_i X_i\right)\label{eq:TFI}\ee 
We choose the antiferromagnetic (AFM) case where $J>0$ since this sign is naturally realized in the Rydberg atom system. 

We can compute the DSF $S^{zz}(k,\omega)$ using the following formula with the retarded Green's function 
\be S^{zz}(k, \omega) = -\frac{1}{\pi}[1 + n_B(\omega)]\mathrm{Im}[G^{\mathrm{ret}}_{zz}(k, \omega)] \label{eq:dsf}\ee 
    where the Bose distribution function at $T=0$ satisfies $n_B(\omega)=0$ for $\omega>0$ and $n_B(\omega)=-1$ for $\omega<0$ such that only positive frequencies contribute.
We can break the process into two steps: ground state preparation and measuring the DSF. 
\subsection{Ground state preparation}
We first prepare the ground state of the TFI $\ket{\psi_0}$. We tested 2 options: the quantum approximate optimization algorithm (QAOA) and an adiabatic-type Ansatz. 

For QAOA, we can use an Ansatz at level $p$ inspired by Trotter evolution like this 
    \be\ket{\psi(\vec{\gamma},\vec{\tau},\vec{\beta})}=\prod_{i=1}^p e^{-i H_{ZZ}[\tau,\beta]}e^{-i H_X[\gamma_i]}\ket{\mathbf{+}}^\otimes\ee
where the parameters $\vec{\gamma},\vec{\tau},\vec{\beta}$ are vectors of length $p$. The qubits are all initialized in the state $\ket{+}=\frac{1}{\sqrt{2}}(\ket{0}+\ket{1})$ which is the uniform superposition of the computational basis states $\ket{0}$ (ground) and $\ket{1}$ (excited). 
The operators $H_X[\gamma]$ and $H_{ZZ}[\tau,\beta]$ depend on variational parameters and approximate the Ising Hamiltonian terms for a chain of length $L$: 
    \bea H_X[\gamma]&=&\gamma\sum_{i=1}^L X_i \nonumber  \\  H_{ZZ}[\tau,\beta] &=&\tau\left[\beta \sum_{i=1}^L Z_i+\frac{C_6}{4} \sum_{i<j} V(i,j)  (Z_i-\mathbb{I})(Z_j-\mathbb{I})\right]\nonumber\\ \label{eq:Ansatz}\eea 
This works well for the critical point, but requires many levels with local detuning to achieve a high fidelity, which is not optimal for Aquila due to the $4\ \mu$s maximum evolution time. 

\begin{figure}
    \centering
    \includegraphics[width=0.45\textwidth]{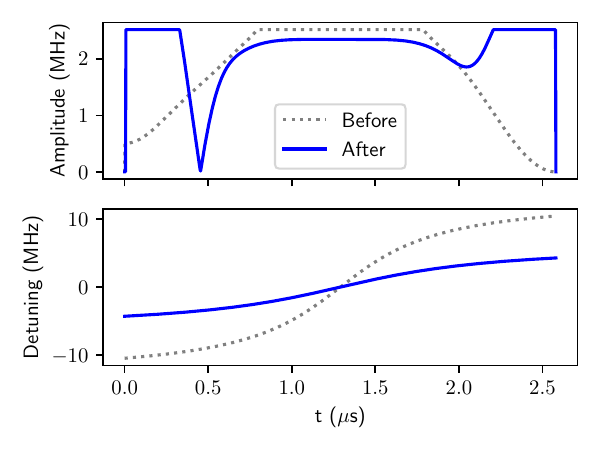}
    \caption{Amplitude (upper) and detuning (lower) pulse shapes for ground state preparation before (solid blue line) and after (dotted grey line) optimization using Pennylane's Nadam optimizer. }
    \label{fig:pulse}
\end{figure}

\begin{figure}
    \centering
    \includegraphics[width=0.45\textwidth]{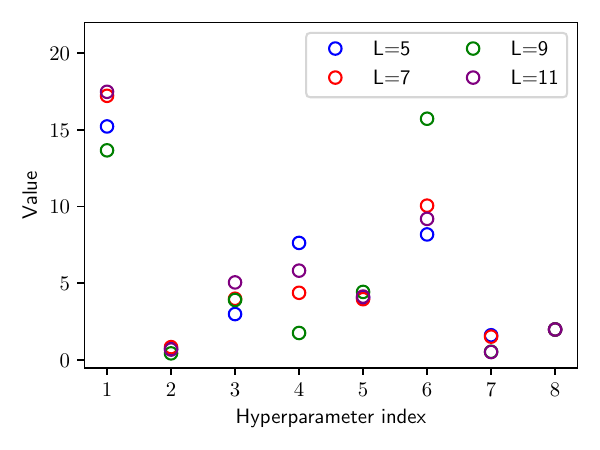}
    \caption{Hyperparameters obtained from optimization of the amplitude and detuning pulses \eqref{eq:ansatz_amplitude},\eqref{eq:ansatz_detuning} using Pennylane's Nadam optimizer for various system sizes $L$.  }
    \label{fig:parameters}
\end{figure}

Instead, we can try an adiabatic type Ansatz, where we can get 94.5\% fidelity for $L=5$ without local detuning. In this method, the qubits are all initialized in the ground state $\ket{0}$ and evolved with Aquila's Hamiltonian \eqref{eq:Aquila}, where the global time-dependent amplitude and detuning pulse forms are given below, and is optimized with respect to a vector of hyperparameters $\vec{p}$. 
\bea \Omega(t,\bm{p}) &=& p_0 \left(1-\left[1-\sin^2\left(\frac{\pi  t}{t_{\mathrm{max}}}\right)\right]^{\frac{p_3}{2}}\right) \nonumber\\
&&+ p_5e^{-5(t-p_4)^4} +p_6e^{-5t^4}  \label{eq:ansatz_amplitude} \eea 
\be \Delta_{\text{global}} (t,\bm{p}) = \frac{2}{\pi} p_2 \tan^{-1}\left(p_1  \left[t-\frac{t_{\mathrm{max}}}{2}\right]\right) \label{eq:ansatz_detuning} \ee 
The pulse shapes for an initial guess of hyperparameters before optimization and after optimization using Pennylane's Nadam optimizer \cite{aws-notebook} are plotted in Fig. \ref{fig:pulse}. 
Most importantly, the hyperparameters appear to have similar behavior for different system sizes. The 8 hyperparameters in the vector $\mathbf{p}$ are plotted for various system sizes $L$ in Fig. \ref{fig:parameters}. 
Thus, further results will use the ground state as prepared with the adiabatic-type Ansatz. 
 \subsection{Measuring DSF}
We would like to measure DSF $S^{zz}(k,\omega)$ \eqref{eq:dsf} of the critical Ising model ground state at temperature $T=0$, which requires measuring the retarded Green's function in Fourier space $G^{\mathrm{ret}}_{zz}(k, \omega)$. 
$G^{\mathrm{ret}}_{zz}(k, \omega)$ can be related to the retarded Green's function in position space and real time $G^{\mathrm{ret}}_{zz}(x,t)$ using the center-site approximation as 
\be G^{\mathrm{ret}}_{zz}(k, \omega)\approx\frac{2\pi\delta}{LT}\sum_{j=1}^Le^{-ik(j-j_c)}\sum_{n=1}^{N}e^{i(\omega+i\eta)t_n}G^{\mathrm{ret}}_{zz}(x,t)\ee 
where $t_n=n\delta$, $T=N\delta$, $L$ is the number of sites, and $j_c$ is the center site. 
Thus, the Fourier transform of the retarded Green's function  $G^{\mathrm{ret}}_{zz}(k, \omega)$ can be obtained by measuring the position/time space Green's function $G^{\mathrm{ret}}_{zz}(x,t)$, which is defined as
\be G^{\mathrm{ret}}_{zz}(i,j,t)=-\frac{i}{2}\braket{ [Z_i(t), Z_j(0)] }_0 \ee 
Note that, in neutron scattering the Green's function is conventionally expressed in spin operators rather than Pauli operators, e.g., $S^z_i=Z_i/2$.

The unequal time correlator can be measured using Raman spectroscopy techniques \cite{Knap_2013}. This topic has been addressed in Ref.\ \cite{Baez_2020}. We first define the spin operator $U_j$ 
\be  U_j=\frac{1}{\sqrt{2}}\left(\mathbf{1}-iZ_j\right)=e^{-i\frac{\pi}{4} Z_j}\ee
Applying this operator to the ground state $\ket{\psi_0}$ and time evolving (under the TFI Hamiltonian ) with unitary $U(t)$ we obtain the state 
\be \ket{\psi}=U(t)U_j\ket{\psi_0}\ee 
The expectation value of the Pauli $Z_i$ operator in this state is 
\be \bra{\psi} Z_i \ket{\psi}=\frac{1}{2} \bra{\psi_0} Z_i(t) \ket{\psi_0}+G^{\mathrm{ret}}_{zz}(i,j,t)+R(i,j,t)\ee 
For the TFI model, the first and third terms vanish due to symmetry arguments because expectation values of odd numbers of Pauli $Z$ operators are equal to 0, and we have 
\be \bra{\psi} Z_i\ket{\psi} = G^{\mathrm{ret}}_{zz}(i,j,t)\ee 

For measuring this on Aquila, we need to apply the local spin operator $U_j$ on the central site, and then time evolve with the Ising Hamiltonian. 

Applying the spin operator $U_{L/2}$ is simple with local detuning because we can do single qubit addressing, so we can do the $Z$ rotation. 

For time evolving under the TFI Hamiltonian for varying $t_n$, we can obtain a very close approximation to the TFI Hamiltonian \eqref{eq:TFI} from the system Hamiltonian \eqref{eq:aquila}.  
Since the interaction $V(i,j)\sim r_{ij}^{-6}$, we can approximate by neglecting terms with $r_{ij}>1$, only considering nearest neighbor interactions. 
\bea  H&\approx&\Omega\sum_{i=1}^n X_i+\sum_{i=2}^{L-1}Z_i \left[\frac{\Delta_i}{2}-\frac{C_6}{2}V(a)\right]\\&&+\sum_{i=1,N} Z_{i} \left[\frac{\Delta_{i}}{2}-\frac{C_6}{4}V(a)\right]+\sum_{i<j}\frac{C_6}{4}  V(a) \left[Z_iZ_j  +  \mathbb{I}\right]  \nonumber  \eea 
Then we can set $\Delta_{2,...,L-1}=C_6 V(a)$ where a is the spacing between atoms. This will make the linear $Z$ terms (except for those at the endpoints) vanish such that we have 
\be H \approx \Omega\sum_{i=1}^n X_i+ \sum_{i=1,N} Z_{i} \left[\frac{\Delta_{i}}{2}-\frac{C_6}{4}V(a)\right] +\frac{C_6}{4}  \sum_{\langle i,j\rangle} V(a)  Z_iZ_j \ee 
where we neglect the $\mathbb{I}$ term, which would only contribute to a global phase. The undesired linear $Z$ terms on the qubits at the endpoints of the chain can be countered by setting $\Delta_{1,N}=C_6V(a)/2$, and we have 
\be H \approx \Omega\sum_{i=1}^n X_i+\frac{C_6 V(a) }{4} \sum_{\langle i,j\rangle}   Z_iZ_j \ee 
Finally, if we set $\Omega=g\frac{C_6 V(a) }{4}$, we have 
\be H \approx J\left(g\sum_{i=1}^n X_i+ \sum_{\langle i,j\rangle}   Z_iZ_j\right) \ ,  \ J= \frac{C_6 V(a) }{4}\label{eq:correspond}\ee 

Measuring the spin operator $Z_i$ on all sites is trivial because measurements are always in the computational/Z basis. Thus, the major source of systematic error in the time evolution is from the next nearest neighbor interactions, since this derivation neglected interactions between qubits with $r_{ij}\geq 2a$. The effect of this systematic error will be discussed in Section \ref{sec:numerical}. 


\section{Error Mitigation}\label{sec:error}
Here we introduce two critical components to our simulations: the noise model and error mitigation. 
\subsection{Noise Model} 
Our noise model builds off of the neutral atom noise model developed by Silv\'erio et. al. in the Pulser package \cite{Silv_rio_2022}. They included SPAM (State Preparation And Measurement), laser, and thermal/effective error. We used similar noise types but specialized to the Aquila processor, and written to be compatible with Pennylane's machine learning modules. Here we will give a brief overview of the noise model and error sources.   
\subsubsection{ SPAM (State Preparation And Measurement) errors}

We first consider the SPAM errors because they can be simulated and corrected for classically, as they correspond to bit flips in the measured bit strings. State preparation errors result from the initial state missing an atom in the register, with probability $\eta$. 
Measurement errors result from an atom in the ground state being falsely measured as the Rydberg state (false positive) with probability $\epsilon$, or an atom in the Rydberg state being falsely measured as the ground state (false negative) with probability $\epsilon'$. 

For this noise type, Aquila has \cite{wurtz2023aquila} 
\be\eta= 0.01^* \ , \ \epsilon= 0.01\, \ \epsilon'= 0.08 \ee


SPAM errors can be corrected for using confusion matrices. 
%
%
%
For SPAM error mitigation with multiple qubits, we can use the matrix inverse method \cite{error-notebook}. 
%



\subsubsection{ Laser noise }  
There are 3 considered contributions from the laser: Doppler damping, finite laser waist, and amplitude fluctuations. 

Doppler damping results from the neutral atoms not being cooled to 0 K, so the laser frequency is Doppler-shifted due to thermal motion, causing a shift in the detuning frequency of the laser. For this noise type, we sample from a Gaussian distribution with standard deviation 
\begin{equation}\sigma_{doppler} = K_{\mathrm{eff}} \sqrt{k_B T/m}\end{equation} to obtain a shift in the detuning at each atom. 

The laser producing the global amplitude has a finite waist with a Gaussian profile such that atoms at the border of the waist feel a slightly lower amplitude than those at the focus. The amplitude of the laser at a distance from the center $r$ is weaker by a factor of $e^{-(\frac{r}{r_{\ell}})^2}$, where $r_{\ell}$ is the radius of the laser. 

There are also amplitude fluctuations from inhomogeneity in the laser. Thus for the amplitude, we sample local amplitude values on each atom by a factor from a Gaussian distribution centered at $e^{-(\frac{r}{r_{\ell}})^2}$ with standard deviation $\sigma_\Omega$. 

For the laser noise, we use the values for parameters given in Table \ref{tab:laser}. 
\begin{table}[]
    \centering
    \begin{tabular}{|c|c|c|c|}\hline
        Parameter & Value & Parameter & Value \\ \hline 
        T & $50 \ \mu$K & $m$ & $1.45\times10^{-25}$ kg\\ \hline
        w & $175 \ \mu$m & $p_z$ &  0.1 \\ \hline 
        $\sigma_\Omega$ & 0.05  & $p_{r0}$ & 0.1 \\ \hline 
        $K_{\mathrm{eff}}$ & $8.7 \ \mu m^{-1}$  & $p_{r1}$ & 0.1 \\ \hline 
    \end{tabular}
    \caption{Values for the parameters used in the laser noise and effective noise channel. }
    \label{tab:laser}
\end{table}
These are slightly over exaggerated values. If we can mitigate this error in numerical simulations with the noise model, then we should be able to reduce the lesser amounts of error when running it on the quantum computer.

\subsubsection{ Effective noise channels } 
Finally, we consider effective noise channels which account for thermal fluctuations and interactions with the environment, which occur because the qubits are not completely cooled to absolute zero and fully isolated from environmental interactions. 
We can model the effects of environmental thermal error using a noise channel with the following Kraus matrices: 
\bea K_0 &=& \sqrt{p_{0}} \left [ \begin{array}{cc}1&0\\0&1 \end{array} \right ], \ K_1 = \sqrt{p_z} \left [ \begin{array}{cc}1&0\\0&-1 \end{array} \right ], \nonumber \\ K_2 &=& \sqrt{p_{r0}} \left [ \begin{array}{cc}1&0\\0&0 \end{array} \right ] , \ K_3 = \sqrt{p_{r0}} \left [ \begin{array}{cc}0&1\\0&0 \end{array} \right ] \nonumber \\ 
K_4 &=& \sqrt{p_{r1}} \left [ \begin{array}{cc}0&0\\1&0 \end{array} \right ] , \ K_5 = \sqrt{p_{r1}} \left [ \begin{array}{cc}0&0\\0&1 \end{array} \right ]  \eea 
where $p_0+p_z+p_{r0} + p_{r1} =1$.
These matrices represent the different single qubit errors that can occur. The coefficients are the probabilities of the different error types, where $p_z$ is the probability of a phase flip (Pauli Z) error, $p_{r0}$ is the probability of a reset to 0 error, and $p_{r1}$ is the probability of a reset to 1 error. These error channels are applied independently on each qubit. 
For the numerical simulations, we average over 5000 independent samples.

\subsection{Error Mitigation} 
Here we propose an error mitigation procedure for both the numerical simulations with the noise model and the experimental results. 
An interesting metric for error mitigation is the survival probability of the computational basis states on qubit $j$, $\mathcal{G}_j$. This is obtained by applying $N_U$ random unitaries sampled from the Haar measure to the prepared state and measuring the resulting bit strings. For $N$ qubits, we have a tensor product unitary $U^{(r)}=u_1^{(r)}\otimes...\otimes u_N^{(r)}   $. We can  construct  the estimator 
\be   \mathcal{G}_j=\frac{12}{5N_U}\sum_{r,s_j}\hat{P}(s_j|u_j^{(r)})P(s_j|u_j^{(r)})-\frac{4}{5}\ee 
where $P(s_j|u_j^{(r)})=|\langle s_j|u_j^{(r)}|0\rangle|^2$ is the theoretical probability of measuring $s_j$ after applying $u_j^{(r)}$ and $\hat{P}(s_j|u_j^{(r)})$ is the noisy probability. 
Haar-sampled unitaries have the form 
\be U(\psi,\theta,\omega)=R_Z(\omega)R_Y(\theta)R_Z(\phi)\ee
where  $\phi$ and $\omega$ are sampled from the uniform distribution $[0,2\pi]$ and $\theta$ is sampled from the probability distribution $P(\theta)=\sin\theta$. 


Then, we can replace the expectation values of the $Z_j$ operator with the noisy estimator 
\be \langle Z_j \rangle = \frac{1}{N_M}\sum_{\mathbf{s_j}}^{N_M} (-1)^{s_j}\rightarrow \mathcal{Z}_j=\frac{1}{N_M}\sum_{\mathbf{s_j}}^{N_M} \left[\frac{\mathcal{G}_j}{2-\mathcal{G}_j}(-1)^{s_j}\right] \ee 
where $N_M$ is the number of shots.



\section{Numerical Results}\label{sec:numerical}
Here we present numerical simulations of the ground state preparation and measurement of the DSF. The numerical simulations are performed using Pennylane's Aquila simulator, which takes into account the specifications of the device and writes a program that can be ran on Aquila. First, we quantify the error introduced by the adiabatic state preparation and approximate time evolution, which we consider to be systematic errors. Then, we run simulations of the hardware error with the noise model described in the previous section, and test the performance of the error mitigation procedure. We also restrict to the critical point where $g=1.0$. 

Before considering the hardware error, we can address the systematic errors resulting from imperfect state preparation and approximate time evolution. 

\begin{figure} 
    \centering
    \includegraphics[width=0.4\textwidth]{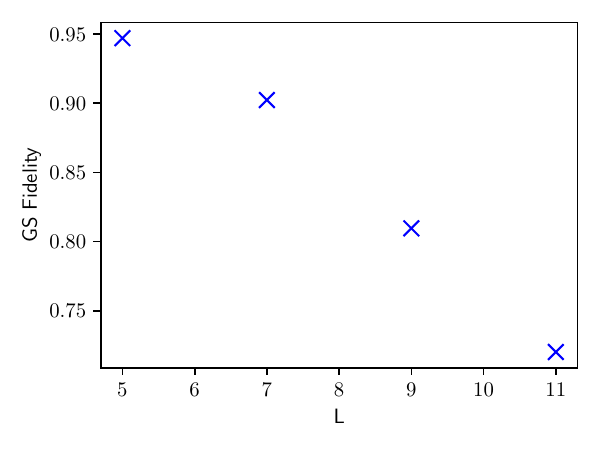} 
    \caption{Ground state fidelity prepared with the adiabatic algorithm for various system size $L$.  } 
    \label{fig:fidelity}
\end{figure}

The fidelity of the adiabatically prepared ground state is shown in Fig.\ \ref{fig:fidelity} for increasing system size $L$,

In addition to the state preparation error, the error in time evolution comes from next-nearest neighbor (NNN) interactions, which we neglected in the derivation of Eq.\ \eqref{eq:correspond}. 

\begin{figure} 
    \centering
    \includegraphics[width=0.49\textwidth]{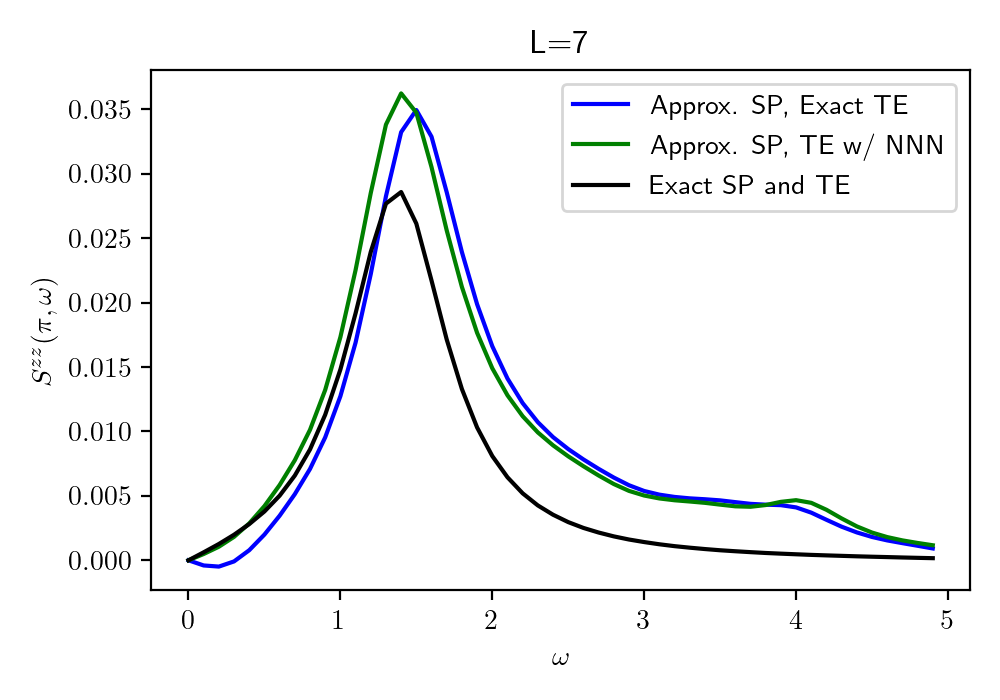} 
    \includegraphics[width=0.49\textwidth]{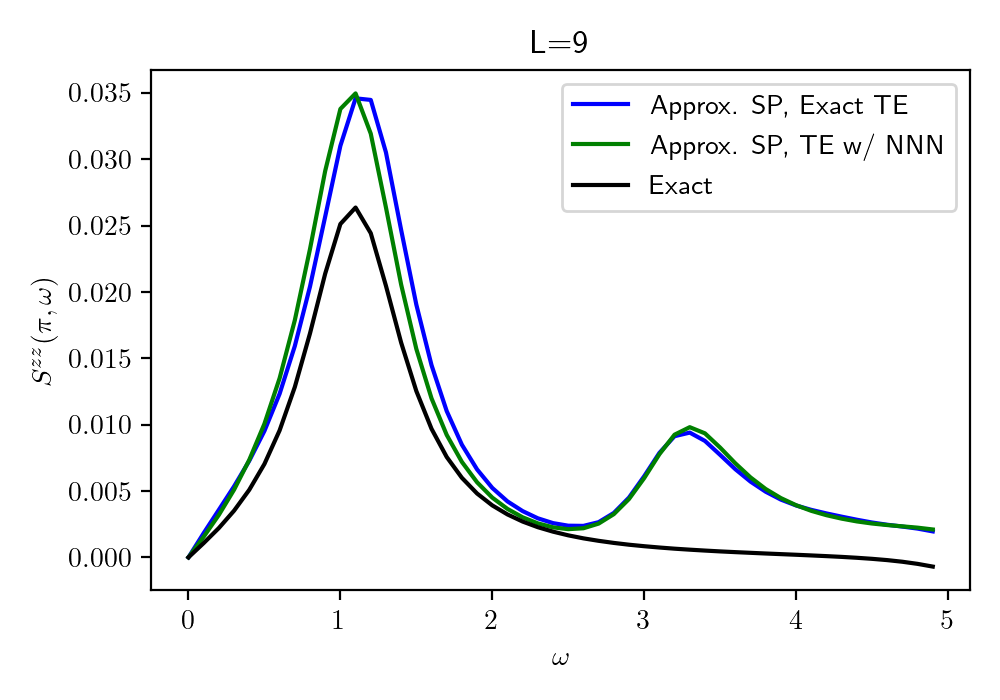} 
    \caption{$S^{zz}(\pi,\omega)$ for 7 and 9  spin lattices. The black line gives the exact value of $S^{zz}(\pi,\omega)$ from matrix calculations, the blue line gives the value from the adiabatic prepared state and exact matrix time evolution, and the green line gives the adiabatic prepared state with approximate time evolution, including the next-nearest neighbor (NNN) interactions. } 
    \label{fig:sxx_num_10_5}
\end{figure}

We can consider the effect of both systematic noise sources  at the critical point $g=1.0$ by calculating the $S^{zz}$. The results in Fig.\ \ref{fig:sxx_num_10_5} are for time evolution over $20~\mu$s. The black line gives the exact result from matrix calculations. The blue line is obtained using the approximate ground state and evolving with exact time evolution, and the green line is obtained by using the approximate time evolution including the NNN interactions. The green and blue lines are closer, indicating that the contribution from the NNN interactions are small for this time scale. However, the error resulting from the ground state preparation error seems to manifest as a bump appearing at larger $\omega$ than the main peak. This indicates that the main source of systematic error is from the state preparation, not time evolution. The main peak magnitude and location does not appear to degrade with increasing system size. 

\begin{figure}
    \centering
    \includegraphics[width=0.45\textwidth]{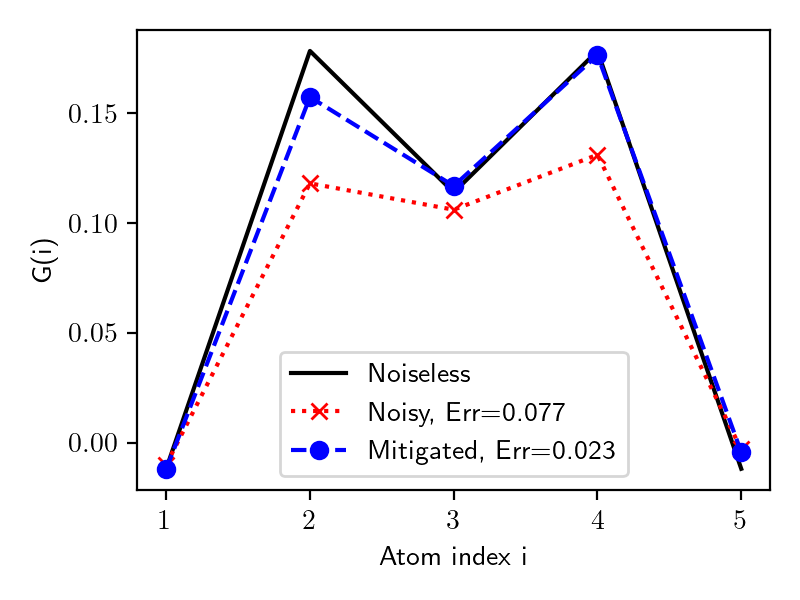}
    \caption{Expectation value of the Green's function for $L=5$ for the approximate state preparation on a noiseless simulator (black), with thermal noise and laser noise (red), and with noise and the error mitigation protocol mentioned above (blue). Note that the approximate state preparation with the adiabatic type algorithm is used in all cases, and $dt=0.005$ in the numerical simulations}
    \label{fig:err_mitigation}
\end{figure}

Now we want to consider the hardware error as simulated using the noise model. First, we can perform a noisy simulation of the state preparation to obtain the comparison shown in Fig. \ref{fig:err_mitigation}. The noiseless Green's function is given as the exact black line. The red line is the results for the measured Green's function, and the blue line shows those results with the error mitigation procedure applied. The sum of deviations from the exact result are given as $\mathrm{Err}$, which is reduced by a factor of 3 using the error mitigation. 

\begin{figure}
    \centering
    \includegraphics[width=0.49\textwidth]{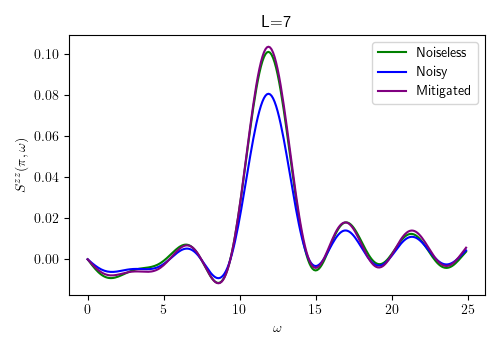}
    \includegraphics[width=0.49\textwidth]{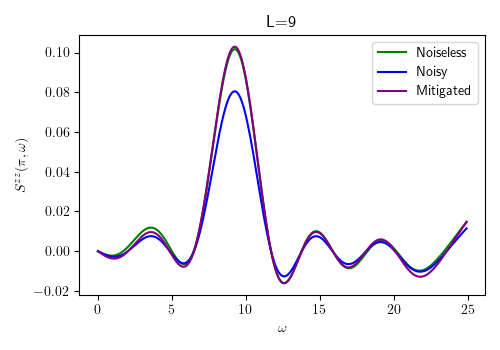}
    \caption{Expectation value of $S^{zz}(\pi,\omega)$ for $L= 7,\ 9$ (a,b) for the noiseless case (green), noisy case with thermal noise and laser noise (blue), and noisy with error mitigation case (purple). In the numerical simulations, $dt=0.005$ and $t_{\mathrm{max}}=20$.}
    \label{fig:err_mitigation_results}
\end{figure}

Now we turn to the time evolution. Rather than measuring the survival probability for the error mitigation at each step, we choose to use the same quantity that captures the bias at the ground state for the error mitigation throughout the time evolution. In Fig.\ \ref{fig:err_mitigation_results}, the results for $S^{zz}(\pi,\omega)$ are shown for the noiseless (green), noisy (blue), and mitigated (purple) simulations. The error mitigation provides a significant advantage, as the results are very close to the exact values. 

\section{Experimental Results}\label{sec:experimental} 
Here we present experimental results on the Aquila processor for the state preparation and DSF computation for the $L=11$ and $L=25$ TFI chain at the critical point where $g=1$. In addition, we compute the QFI density experimentally. We also discuss the constraints introduced by the current hardware and how the procedure can be improved. 

\begin{figure} 
    \centering
    \includegraphics[width=0.49\textwidth]{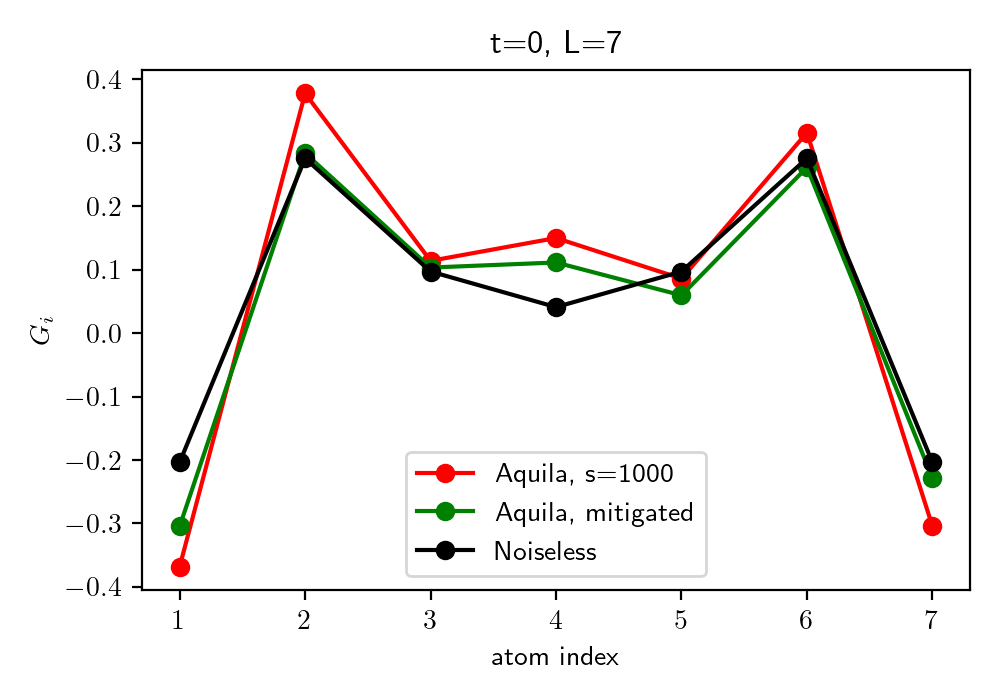}
    \caption{Measured spatial Green's function $G_i$ for $L=7$ for the ground state as prepared with the adiabatic algorithm. The black line gives the noiseless result, and the red line gives the result as measured on the Aquila processor. Then, the green line gives the result after applying the error mitigation scheme where the survival probabilities are computed using the noisy simulator. }
    \label{fig:exp_1}
\end{figure}

Expectation values of the ground state as prepared experimentally for $L=7$ with and without error mitigation are given in Fig. \ref{fig:exp_1}. This is prepared without local detuning. The error mitigated results use the survival probability values as measured on the noisy simulator. The error mitigated results are generally closer to the exact values. 


Now we can do the central qubit rotation and time evolution using  local detuning. Note that, due to hardware limitations on Aquila, the current implementation of local detuning uses a time dependent function $\Delta_{\mathrm{local}}$ and detuning pattern $\mathbf{s}=\{s_i\}$ with $s_i\in [0,1]$ as 
\be \sum_i \Delta_i Z_i\rightarrow \Delta_{\mathrm{local}}\sum_i s_i Z_i\ee 
The coefficients $s_i$ can only be chosen once and not varied in time, so if we set $s_{L/2}=1$ and the rest $s_i=0$ to perform the local rotation, we cannot choose another local detuning pattern for the time evolution where $s_{0,N}=1$ and the rest $s_i=0$. For our experiments, we choose to use the local detuning for the local rotation and only use global detuning for the time evolution. 

\begin{figure}
    \centering
    \includegraphics[width=0.49\textwidth]{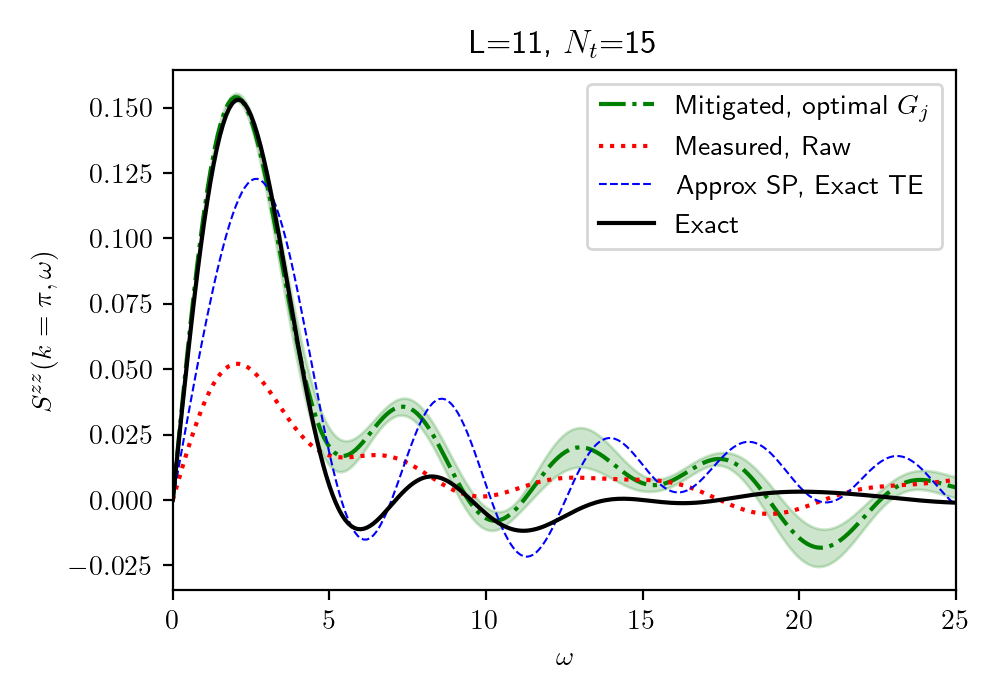}
    \caption{$S(k,\omega)$ results measured on QuEra's Aquila device for $15$ time steps. The black solid line denotes the result with exact state preparation (SP) and time evolution (TE). The blue dashed line denotes the result with approximate state preparation (from the adiabatic Ansatz) and exact time evolution. The red dotted line denotes the raw measured data from Aquila. The green dash-dot line denotes the mitigated experimental results, with an optimized $G_j$ value, where the shading denotes $3\sigma$ standard deviation. There is also a correction for finite time step in the quantum program. The next step is to measure this experimentally.  }
    \label{fig:exp_TD_11}
\end{figure}

The results for $S^{zz}(k,\omega)$ as measured using simulation and experimentally using Aquila for $L=11$ are given in Fig.\ \ref{fig:exp_TD_11}. The results from the raw data measured on Aquila are dampened and have smaller peaks, and the error mitigation increases the amplitude to something very close to the ideal value. Also shown is the result where exact time evolution is applied to the approximate (but noiseless) ground state. The error mitigation procedure appears to diminish effects from both the approximate state preparation and the hardware noise. We attribute the small negative values of the DSF to a combination of the center site approximation and the finite number of time steps $N_t$. 
We now consider a larger system. For $L=25$, we use the large $L$ parameters as shown in Fig.\ \ref{fig:parameters} for preparing the ground state on Aquila. 

\begin{figure}
    \centering
    \includegraphics[width=0.49\textwidth]{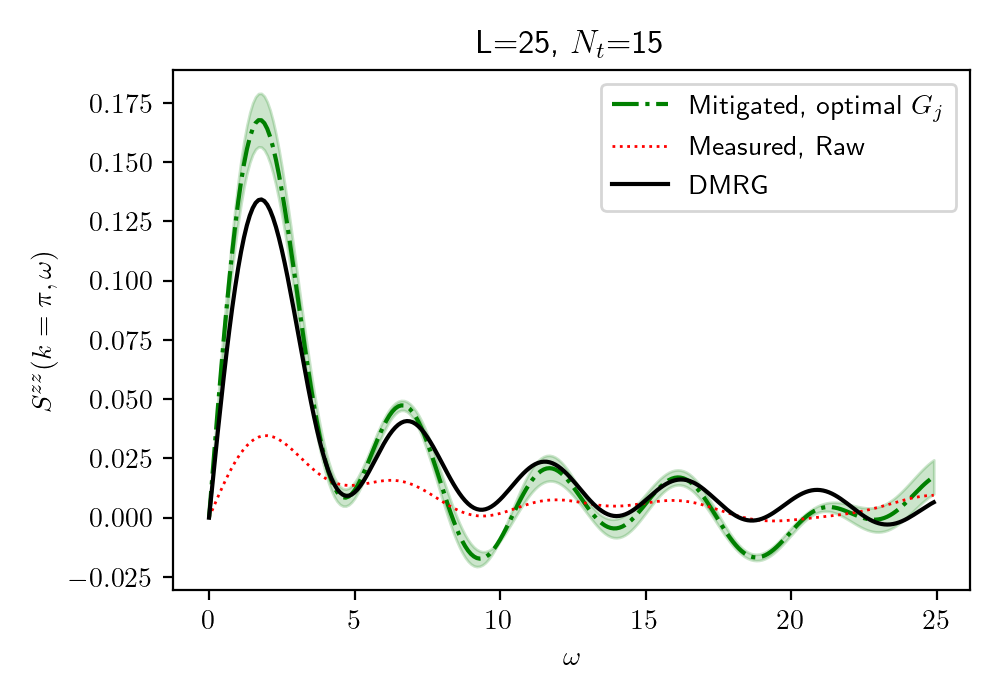}
    \caption{$S(k,\omega)$ results measured on QuEra's Aquila device for $14$ time steps. The black solid line denotes the result from DMRG. The red dotted line denotes the raw measured data from Aquila. The green dash-dot line denotes the mitigated experimental results, where the shading denotes $3\sigma$ standard deviation. There is also a correction for finite time step in the quantum program. The next step is to measure this experimentally.  }
    \label{fig:exp_TD_25_1}
\end{figure}

Tensor networks are used to obtain an approximate classical result for comparison. The ground state of the TFI is obtained using the density matrix renormalization group (DMRG) algorithm from the TeNPy package \cite{tenpy}. We find convergence for $L=25$ with bond dimension $\chi=400$. Then, a local matrix product operator (MPO) for the $Z_i$ is applied and time evolution is performed over the same number of time steps using the MPO evolution algorithm from \cite{Zaletel_2015}. The DMRG results and experimental results with and without error mitigation are shown in Fig.\ \ref{fig:exp_TD_25_1}. 

As in the $L=11$ result, the values for $S^{zz}(\pi,\omega)$ using the raw data are dampened compared to the exact matrix result. The mitigated results slightly overshoot the DMRG result, but are closer. 

Now we turn to computing the quantum Fisher information (QFI). The QFI density $f_Q$ is given as \cite{Hauke_2016}
\be f_Q(k,T)=\frac{4}{\pi}\int_0^\infty d(\hbar \omega)\tanh\left(\frac{\hbar\omega}{2k_BT}\right)\chi_{zz}''(k,\hbar\omega,T)\label{eq:QFI}\ee 
where $\chi_{zz}''$ is the dynamic susceptibility, related to the frequency-symmetrized DSF $\widetilde{S}(k,\omega)\equiv S(k,\omega)+S(k,-\omega)$ by the fluctuation dissipation theorem as 
\be \chi_{zz}''(k,\hbar\omega,T)=\tanh\left(\frac{\hbar\omega}{2k_BT}\right)\widetilde{S}^{zz}(k,\hbar\omega)\ee 
giving the expression 
\be f_Q(k,T)=\frac{4}{\pi}\int_0^\infty d(\hbar \omega)\tanh^2\left(\frac{\hbar\omega}{2k_BT}\right)\widetilde{S}^{zz}(k,\hbar\omega)\ee 
For $T=0$, we have $\tanh^2\left(\frac{\hbar\omega}{2k_BT}\right){2k_BT}=1$ and $\widetilde{S}(k,\omega)=S(k,\omega)$ such that
\be f_Q(k,T=0)=\frac{4}{\pi}\int_0^\infty d(\hbar \omega)S^{zz}(k,\hbar\omega)\label{eq:QFI2}\ee 
To evaluate the QFI $S^{zz}(k,\omega)$ must be obtained in absolute units, which is guaranteed if it is normalized per site by the sum rule \cite{Laurell_2021}
\be \sum_{\alpha=x,y,z}\int_{-\infty}^\infty d(\hbar \omega)\int_0^{2\pi}dk S^{\alpha\alpha}(k,\hbar\omega)=3.\label{eq:norm}\ee
Note that, in neutron scattering the sum rule is conventionally expressed in spin operators rather than Pauli operators, e.g., $S^z_i=Z_i/2$, 
making the right-hand side $S(S+1)$, with spin $S=1/2$ in our case.

%
For the normalization \eqref{eq:norm}, we use the normalization constant obtained from exact matrix calculations ($L=11$) or DMRG ($L=25$). Additionally, we used a cutoff frequency value $\omega_{\mathrm{max}}=25$ for the integration over $\omega$ in Eq. \eqref{eq:QFI2}. We used a window of equivalent size for the integration in the normalization \eqref{eq:norm}. 

Finally, we define the normalized QFI density \cite{scheie2024tutorialextractingentanglementsignatures} 
\be \Tilde{f}_Q=\frac{f_Q}{\left( h_\mathrm{max}-h_\mathrm{min}\right)^2} = \frac{f_Q}{4},\ee 
where $h_\mathrm{max}=1$ ($h_\mathrm{min}=-1$) is the maximal (minimal) eigenvalue of the probed local $Z_i$ operator. 
We can compare our results with the analytic value obtained by measuring the lower bounds of the QFI $F_n$ until they converge \cite{Vitale_2024}, using the formula 
\be F_n=2\sum_{q=0}^n\begin{pmatrix} n+1 \\ q+1 \end{pmatrix} (-1)^q\sum_{m=0}^{q+2}C_m^{(q)}\mathrm{Tr} (\rho^{q+2-m}A\rho^mA)\label{eq:estimator}\ee 
where $C_m^{(q)}=(\begin{smallmatrix} q \\ m \end{smallmatrix})-2(\begin{smallmatrix} q \\ m-1 \end{smallmatrix})+(\begin{smallmatrix} q \\ m -2\end{smallmatrix})$, given in terms of binomial coefficients,  and $A=\sum_iZ_i$. We find that for the $L=11$ system, the bound converged by $n=1$. 

\begin{figure}
    \centering
    \includegraphics[width=0.45\textwidth]{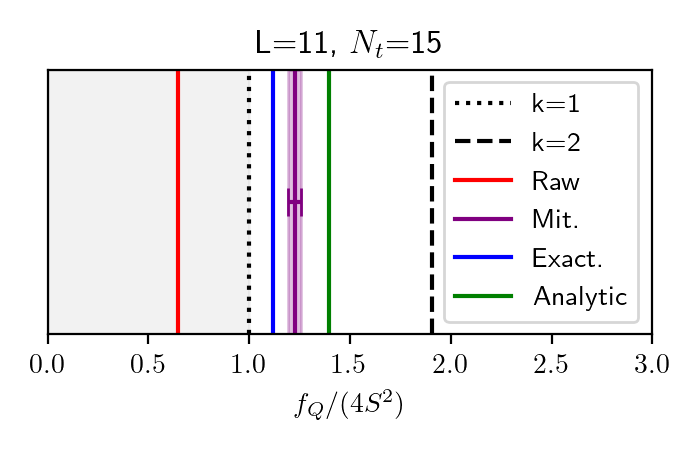}
    \includegraphics[width=0.45\textwidth]{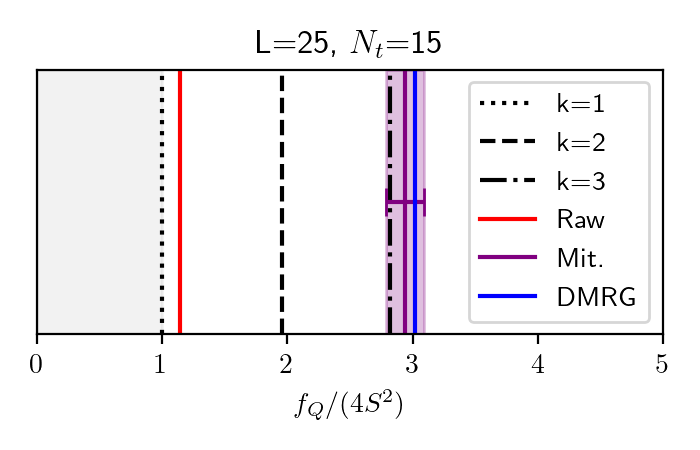}
    \caption{Measured values for the normalized QFI density $\Tilde{f}_Q$ for the raw experimental data (red), mitigated experimental data (purple), and exact matrix/DMRG result (blue) for $L=11$ (upper) and $L=25$ (lower) with $N_{t}=15$. For $L=11$, we also give an analytic value from Eq. \eqref{eq:estimator} (green). The purple shading around the mitigated values $1\sigma$ standard deviation. The black lines show the QFI density threshold for $k+1$-entanglement depth for $k=1$ (dotted), $k=2$ (dashed), and $k=3$ (dash-dotted). The grey shaded region denotes where $f_Q$ indicates a separable state. }
    \label{fig:qfi_calc}
\end{figure}

The results for the normalized QFI density $\Tilde{f}_Q$ for the $L=11$ and $L=25$ systems from numerical simulations and experiment are shown in Fig. \ref{fig:qfi_calc}. Additionally, we give the threshold values for $\Tilde{f}_Q$ for $k+1$ entanglement depth \cite{Vitale_2024} in a system of size $N$ given as 
\be \Tilde{f}^k_Q(N)=\frac{\lfloor N/k \rfloor k^2 + (N- \lfloor N/k \rfloor k)^2}{N}\ee 
where $\lfloor \cdot \rfloor$ is the floor function. For $L=11$, we find that the raw measured values do not give a normalized QFI density indicative of entanglement, but with the error mitigation, we have at least bipartite (depth 2) entanglement. The analytic normalized QFI density value from Eq. \eqref{eq:estimator} is higher than the QFI density from the raw, mitigated, and exact matrix calculation values. For $L=25$, we find a normalized QFI density indicative of bipartite entanglement in the raw measured values, and a value indicative of at least 4-partite entanglement for the mitigated and DMRG values. The mitigated value is within one standard deviation of the DMRG value. We would like to emphasize that we can observe at least bipartite entanglement with the raw measured data before applying error mitigation.  

In the future, we hope to extend the procedure to measure $S^{xx}(k,\omega)$ and $S^{yy}(k,\omega)$ values. The current limitation is the maximum evolution time of 4 $\mu$s on the Aquila processor, since we could approximate the $X$ and $Y$ rotations and measurements using additional pulses, but this would diminish the amount of time left for time evolution. 

\section{Conclusion}\label{sec:conclusion} 

Neutron scattering experiments, which provide insights into the microscopic structure and dynamics of materials, are notoriously difficult to simulate classically, especially for systems with strong correlations, long-range interactions, and large system sizes. Quantum processors, particularly analog quantum simulators like those based on Rydberg atom arrays or trapped ions, have the potential to transform this field by simulating the underlying quantum dynamics directly.
In particular, the ability of quantum simulators to compute dynamical structure factors—a key quantity measured in inelastic scattering experiments—offers a significant advantage. Classical simulations of these quantities are computationally expensive due to the difficulty of tracking the time evolution and entanglement growth of quantum systems with many degrees of freedom. However, analog quantum processors, which naturally evolve under a Hamiltonian closely mimicking the one in a real material, could efficiently simulate the time-dependent correlation functions central to neutron scattering studies. This capability would allow us to probe the dynamic properties of quantum materials, including excitations, phase transitions, and emergent phenomena like spin liquids, in a way that is not feasible on classical computers. Indeed, neutron scattering experiments on quantum magnets have been shown to be highly accurate making them ideal platforms for the experimental verification of quantum computations beyond the reach of classical hardware benefiting a wide range of quantum applications.

We demonstrated the feasibility of using analog quantum processors like Aquila for simulating neutron scattering experiments, specifically the dynamic structure factor of quantum spin chains. Error mitigation techniques significantly improved the fidelity of results, and future work will focus on extending these simulations to more complex systems and improving entanglement detection. Specifically, we are interested in extending our work to 2+ dimensions, where classical methods like tensor networks encounter more difficulty. Additionally, the procedure can be extended to study properties of out of equilibrium systems and thermal states prepared using the quantum imaginary time evolution (QITE) algorithm \cite{bangar2024continuousvariablequantumboltzmannmachine,Wang_2023}. 

In the near future, analog quantum processors could also simulate strongly correlated materials that are otherwise intractable, providing insights into systems such as high-temperature superconductors, frustrated magnets, and quantum spin liquids. These systems are of immense interest in condensed matter physics but pose extreme computational challenges. Quantum processors can offer an experimental pathway to explore these materials, uncovering new physical insights and enabling direct comparisons with neutron scattering data.
Furthermore, as quantum simulators advance, there will be opportunities to incorporate error mitigation techniques and improve the fidelity of analog quantum simulations, pushing the boundaries of accessible system sizes and interaction complexities. This progress could ultimately lead to simulations that match or exceed the precision of classical methods, making analog quantum processors indispensable tools for studying the quantum properties of materials.

In the long term, simulating neutron scattering and other spectroscopic techniques on quantum processors could revolutionize the design of quantum materials, enabling the prediction of new phases of matter and guiding the development of next-generation technologies like quantum sensors, superconductors, and spintronic devices. By leveraging the full potential of quantum simulation, we can explore materials with unprecedented accuracy, opening up new pathways for discovery across multiple fields of science and engineering.

\acknowledgments
Research supported by the National Science Foundation under awards DGE-2152168 and CNS-2244512 and the Department of Energy under awards DE-SC0024325 and DE-SC0024328.
The work by D.A.T. is supported by the Quantum Science Center (QSC), a National Quantum Information Science Research Center of the U.S. Department of Energy (DOE).
A portion of the computation for this work was performed on the University of Tennessee Infrastructure for Scientific Applications and Advanced Computing (ISAAC) computational resources. The Aquila processor was accessed through Amazon Braket, and we acknowledge support from the AWS Cloud Credit for Research program. 

%

 
\end{document}